\documentclass[10pt,journal]{IEEEtran}

\usepackage{graphicx,times,cite,amsmath, amssymb, amsthm, mathrsfs, epsfig, array, subfigure}

\usepackage{multirow}

\usepackage[colorlinks, linkcolor=black, anchorcolor=black, citecolor=black, urlcolor=black]{hyperref}
\usepackage{cite}

\usepackage{color}

\allowdisplaybreaks[2]
\newcommand{\figwidthS}{2.3in}
\newcommand{\figwidth}{\linewidth}
\newcommand{\figwidthW}{0.6\linewidth}
\newcommand{\fref}[1]{Fig.~\ref{#1}}

\usepackage[acronym]{glossaries} 
\newacronym{iid}{i.i.d.}{independent and identically distributed}
\newacronym{SNR}{SNR}{signal-to-noise ratio}
\newacronym{SINR}{SINR}{signal-to-interference-plus-noise ratio}
\newacronym{AWGN}{AWGN}{additive white Gaussian noise}
\newacronym{MAC}{MAC}{multiple access channel}
\newacronym{BF}{BF}{bit-flipping}
\newacronym{IoT}{IoT}{Internet-of-Things}
\newacronym{AI}{AI}{artificial intelligence}
\newacronym{PDF}{PDF}{probability density function}
\newacronym{CF}{CF}{compress-and-forward}
\newacronym{DF}{DF}{decode-and-forward}
\newacronym{AF}{AF}{amplify-and-forward}
\newacronym{LF}{LF}{lossy-forward}
\newacronym{CEO}{CEO}{chief executive officer}
\newacronym{NOMA}{NOMA}{non-orthogonal multiple access}
\newacronym{SF}{SF}{semantic-forward}
\newacronym{LDPC}{LDPC}{low-density parity-check}
\newacronym{RD}{R-D}{relay-destination}
\newacronym{SR}{S-R}{source-relay}
\newacronym{SD}{S-D}{source-destination}
\newacronym{ED}{ED}{Euclidean distance}
\newacronym{NN}{NN}{neural network}
\newacronym{LoS}{LoS}{line-of-sight}

\begin{document}
\title{\huge Semantic-Forward Relaying: A Novel Framework Towards \\ 6G Cooperative Communications}
\newcommand{\authorsize}{}
\author{\authorsize
Wensheng Lin,~\IEEEmembership{\authorsize Member,~IEEE},
Yuna Yan,
Lixin Li,~\IEEEmembership{\authorsize Member,~IEEE}, 

Zhu Han,~\IEEEmembership{\authorsize Fellow,~IEEE}
and Tad Matsumoto,~\IEEEmembership{\authorsize Life Fellow,~IEEE}
 
\thanks{
This letter has been accepted for publication in IEEE Communications Letters with DOI: \href{https://doi.org/10.1109/LCOMM.2024.3352916}{10.1109/LCOMM.2024.3352916}.

Corresponding author: Lixin Li.

W. Lin, Y. Yan and L. Li are with the School of Electronics and Information, Northwestern Polytechnical University, Xi'an, Shaanxi 710129, China (e-mail: linwest@nwpu.edu.cn; yanyuna@mail.nwpu.edu.cn; lilixin@nwpu.edu.cn).

Z. Han is with the Department of Electrical and Computer Engineering at the University of Houston, Houston, TX 77004 USA, and also with the Department of Computer Science and Engineering, Kyung Hee University, Seoul, South Korea, 446-701 (e-mail: hanzhu22@gmail.com). 

T. Matsumoto is an Invited Professor at IMT-Atlantique, France.  He is also Professor Emeritus of Japan Advanced Institute of Science Technology, Ishikawa 923-1292, Japan, and University of Oulu, Finland (e-mail: matumoto@jaist.ac.jp).
}
}
\markboth{}{}
\maketitle

\begin{abstract}
This letter proposes a novel relaying framework, semantic-forward (SF), for cooperative communications towards the sixth-generation (6G) wireless networks. 
The SF relay extracts and transmits the semantic features, which reduces forwarding payload, and also improves the network robustness against intra-link errors.
Based on the theoretical basis for cooperative communications with side information and the turbo principle, we design a joint source-channel coding algorithm to iteratively exchange the extrinsic information for enhancing the decoding gains at the destination.
Surprisingly, simulation results indicate that even in bad channel conditions, SF relaying can still effectively improve the recovered information quality.

\end{abstract}

\section{Introduction}
Cooperative communications are acknowledged schemes to improve the transmission quality.
One of the most important categories of cooperative communications is in the form of relaying.
Although relaying requires extra energy and time slot, it is an effective solution when the path-loss of the direct link is very large.
On the other hand, with the research trend towards the sixth-generation (6G) wireless networks \cite{Wang20236G}, various transmission technologies have been invented, among which semantic communications \cite{Yang2023Semantic} are considered to have a great potential in media transmissions.
The semantic encoder extracts the semantic features for transmissions \cite{Zhang2023Deep}, while the semantic decoder works in a similar way to the generative \gls{AI} \cite{Hao2024SCAI}.
Inspired by the principle of semantic communications, this letter proposes a novel framework, \gls{SF} relaying, for cooperative communications. 

There have been already diverse relaying schemes in the literature \cite{Kramer2005Cooperative}.
One simple relaying scheme is \gls{AF}, in which the relay directly amplifies the signals received from the source and then forwards to the destination.
In 1979, Cover and El Gamal \cite{Cover1979Capacity} established the fundamental theorems of relaying systems, and proposed the \gls{DF} and \gls{CF} schemes.
In the DF scheme, the relay decodes the received signals at the first step, and then the recovered information sequence is forwarded or discarded, respectively, depending on the recovery is error-free or not.
In the CF scheme, the relay quantizes and compresses its received signals into the relay information to be transmitted to the destination.
Beyond DF, \gls{LF} \cite{Lin2019Lossy} was proposed to overcome the drawback of DF, where the communication resources are completely wasted once errors occur in the relay information.
In the LF scheme, the relay always forwards the relay information to the destination regardless of whether or not intra-link error is detected at the relay.
At the destination, a joint decoder recovers the source information with the help of the relayed information, based on the principle of correlated sources transmission.

Nevertheless, the previous relaying schemes are designed for general types of information, which do not exploit the features of information to improve the information efficiency.
By adopting the semantic communications, the system adaptively exploits diverse types of information.
Hence, we aims at designing a relaying systems where the relay forwards the semantic information to the destination, i.e., SF relaying, so that the destination can utilize the semantic information to help recovering the source information.

The terminology of SF has been used for the first time in \cite{Luo2022Autoencoder}, up to the authors' maximum knowledge, where the source transmits semantic information to the relay, and the relay translates and forwards the processed semantic information to the destination.
However, there is no direct source-destination link in \cite{Luo2022Autoencoder}.

Different from the relay-assisted semantic communications in \cite{Luo2022Autoencoder}, this work proposes a semantic-assisted relaying system.
In our proposed SF relaying system, the relay reconstructs the information received from the source at the first step.
Then, in spite of whether or not the reconstruction error-free, the relay extracts the semantic information and sends it to the destination.
The semantic coding achieves robustness of the relaying system against errors, and hence can reduce the payload in the \gls{RD} link.
At the destination, a joint decoder performs iterative decoding utilizing the Turbo principle \cite{berrou1996near} that exchanges the extrinsic information between the lossy information of the \gls{SD} link and the semantic information of the \gls{RD} link.
In this way, the SF relaying can help for the lossless recovery of the original information at the destinations, even in bad channel conditions.
With the pre-trained semantic encoder/decoder, the SF relaying can reduce the payload of the R-D link in practical systems.
The contributions of this letter are summarized as follows:
\begin{itemize}
\item We propose a novel relaying framework, i.e., SF relaying, which adopts semantic communications at the relay and the destination to reduce the payload of the \gls{RD} link. 

\item We design a joint source-channel coding algorithm for SF relaying systems, where the destination can losslessly recover the source information with the assistance of the semantic information received from the relay.

\item We conduct a series of simulations with image transmissions to evaluate the performance of SF relaying. 
The simulation results demonstrate that SF relaying systems can exploit the semantic information to reduce the \gls{ED} and improve the image quality.
\end{itemize}

\emph{Notation.}
Capital letters $X,Y,V,U$ denote the random variables for constructing information sequences.
$M$ represents the codeword satisfying the link rate constraint $R$.


\section{The Principle of Semantic-Forward Relaying}\label{sec:Principle}
This section explains the principle of SF relaying, and then analyzes the rate constraints on the rates for losslessly recovering the source information.

\subsection{Framework Structure}
\begin{figure}[!t]
\centering \includegraphics[width=\figwidth]{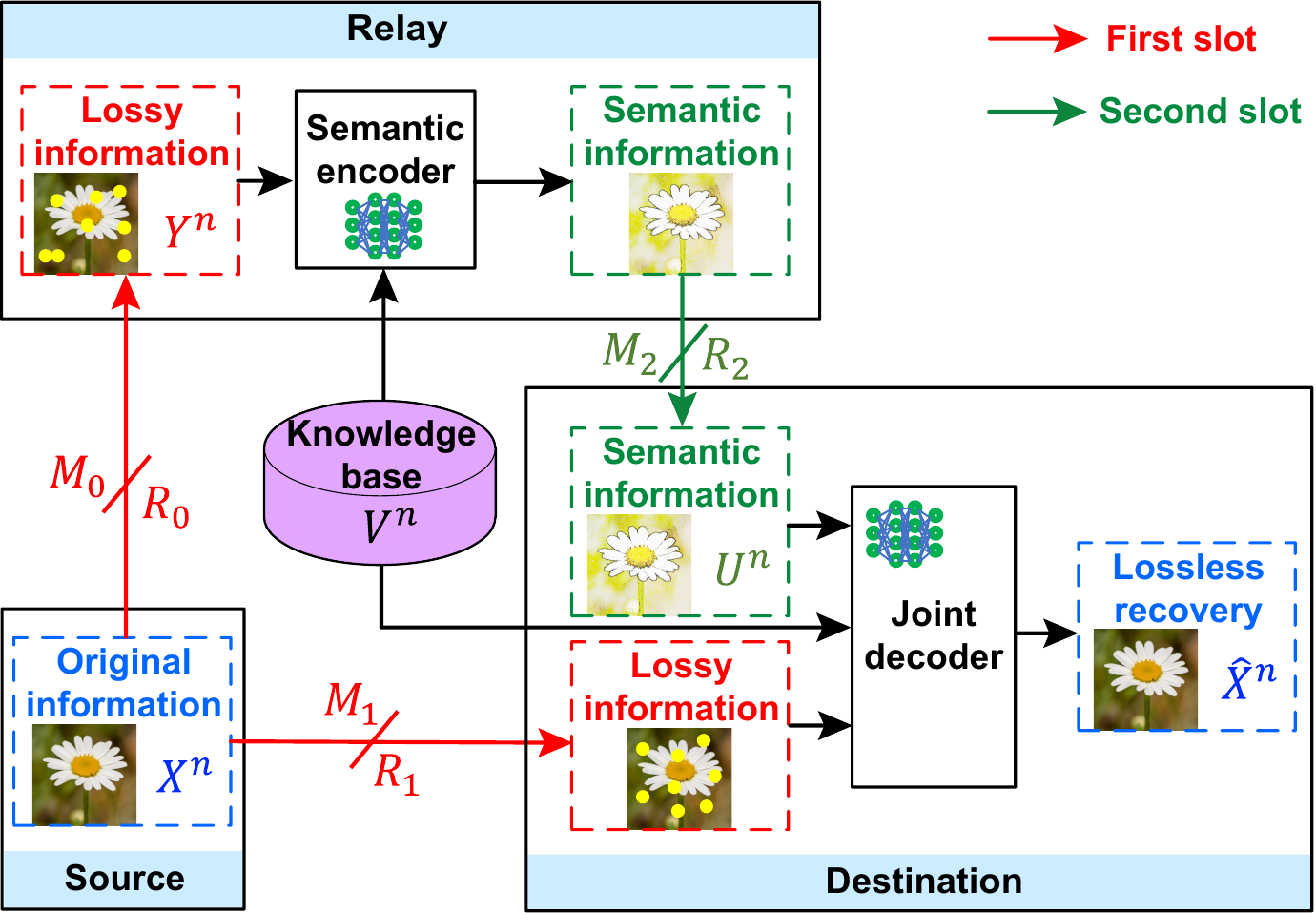}
\caption{The principle of SF Relaying.}
\label{fig:SF}
\end{figure}

As illustrated in \fref{fig:SF}, the SF relaying system contains one source, one relay and one destination.
The source is nothing specific than a common source in conventional relaying systems.
It broadcasts the original information to the relay and the destination at the first time slot.
For semantic communications, the media type of the original information could be image, audio, video, text, etc., and combinations of multimedia.

After receiving the signals from the source, the relay reconstructs the source information as high fidelity as possible.
However, due to the unavoidable channel fading, the reconstructed information may not be lossless.
It is noticed that in semantic communications, the transmitted information is the extracted features of the original information.
This indicates that semantic coding is robust against the noise to a certain extent.
Hence, the relay utilizes a semantic encoder to extract the semantic information based on a common knowledge base shared with the destination.
Then, the semantic information is sent to the destination at the second slot.

Once the destination receives all the signals sent from the source and the relay, it initiates the process for reconstructing the original information by a joint decoder with the assistance of the semantic information, based on the information theoretic principle of cooperative communications with correlated sources.
If the channel conditions of the three links can satisfy the lossless reconstruction requirements at the destination, the destination can recover the original information loslessly.

\emph{Remark 1}: The source can broadcast the semantic information instead of the original information.
However, the recovery at the destination is not easy to be lossless in practical, because it equivalently requires the semantic decoder to losslessly recover the original information by only utilizing the semantic encoded information.

\emph{Remark 2}: In general, the semantic communications aims to transmit much less payload than the original information, while maintaining a relatively high quality of the recovered information.
Thus, given the total transmit energy constant, the transmit energy per bit of the \gls{RD} link can be higher than the \gls{SD} link.

\subsection{Theoretical System Model}

For theoretical analysis, we can rely on the system model illustrated in \fref{fig:SF} in either symbol-wise or bit-wise.
We assume the system to be bit-wise hereafter for simplicity  without loss of generality.
$X^n$ denotes the original information sequence, with $n$ being the information sequence length.
$Y^n$ is the lossy information sequence recovered at the relay.
Hence, $Y$ can be represented by $Y=X \oplus E$, where $E \sim \mathrm{Bern}(\rho)$ is the corrupting error with $\rho$ being the crossover probability between $X$ and $Y$.
Due to the rate constraint $R_1$ supported by the channel capacity on the \gls{SD} link, $X^n$ may not be losslessly transmitted to the destination. 
Therefore, the transmission on the \gls{SD} link can be equivalently regarded as the lossy compression from  $X^n$ to codeword $M_1$ by encoder (ENC) 1.
Similarly, $Y^n$ is also equivalently encoded into codeword $M_2$ by ENC 2 to satisfy rate constraint $R_2$.
$V^n$ represents the side information provided by the semantic knowledge, which is commonly shared by the relay and the destination.
$U^n$ stands for the lossy compressed version of $Y^n$ reconstructed from codeword $M_2$.
Finally, the joint decoder (DEC) reconstructs $\hat{X}^n$ based on codewords $M_1$ and $M_2$ with the assistance of the side information $V^n$.
In this letter, we aim at the lossless recovery of $X^n$, while the recovery $\hat{X}^n$ can be either lossless or lossy in general systems.

\subsection{Achievable Rate Analysis}
Form the Shannon theory, the rate constraint on the \gls{SR} link is
\begin{align}
R_0 \ge I(X;Y). \label{eq:R0}
\end{align}

For the rate constraints on the \gls{SD} and \gls{RD} links, we need to consider from the viewpoint of the joint DEC for losslessly reconstructing $\hat{X}^n$.
First, if without the side information $V^n$, the coding rates need to satisfy the conditions for lossless source coding with a helper \cite[Theorem 10.2]{el2011network}, as
\begin{align}
R_1 &\ge H(X|U),\\
R_2 &\ge I(Y;U).
\end{align}
Then, if the side information is available at both ENC 2 and joint DEC, the \gls{RD} link will be reduced by the conditional version of the mutual information without side information \cite[Eq. (11.2)]{el2011network}, i.e., 
\begin{align}
R_2 &\ge I(Y;U|V). \label{eq:R2}
\end{align}
Furthermore, considering that the side information is available at joint DEC but unavailable at ENC 1, the structure is a Wyner-Ziv problem \cite{wyner1976rate} because the side information provided by the semantic knowledge is noncausal.
Hence, the rate is reduced by condition on $V$, as
\begin{align}
R_1 &\ge H(X|U,V). \label{eq:R1}
\end{align}

In summary, the achievable rate region for lossless transmissions with SF relaying is the combination of \eqref{eq:R0}, \eqref{eq:R2} and \eqref{eq:R1}.
The difference between SF and other relaying systems can be observed from \eqref{eq:R2} and \eqref{eq:R1}.
By introducing a common knowledge base, the required rates in the S-D and R-D links can both be further decreased by the side information $V$.
If the Shannon rate limit is found to be lower than the link payload, we update the  knowledge base to further reduce the rate requirements.

Besides the conventional information theory, we can also apply the metrics of semantic information theory  \cite{Xin2023Semantic}, such as semantic entropy and semantic distortion, to evaluate the semantic performance.

Note that when designing the optimal encoder and decoder, the distributions of the variables need to be known, which is internally estimated by the deep neural network (DNN). 

\section{Joint Source-Channel Coding Design}\label{sec:Coding}

\begin{figure}[!t]
\centering \includegraphics[width=\figwidth]{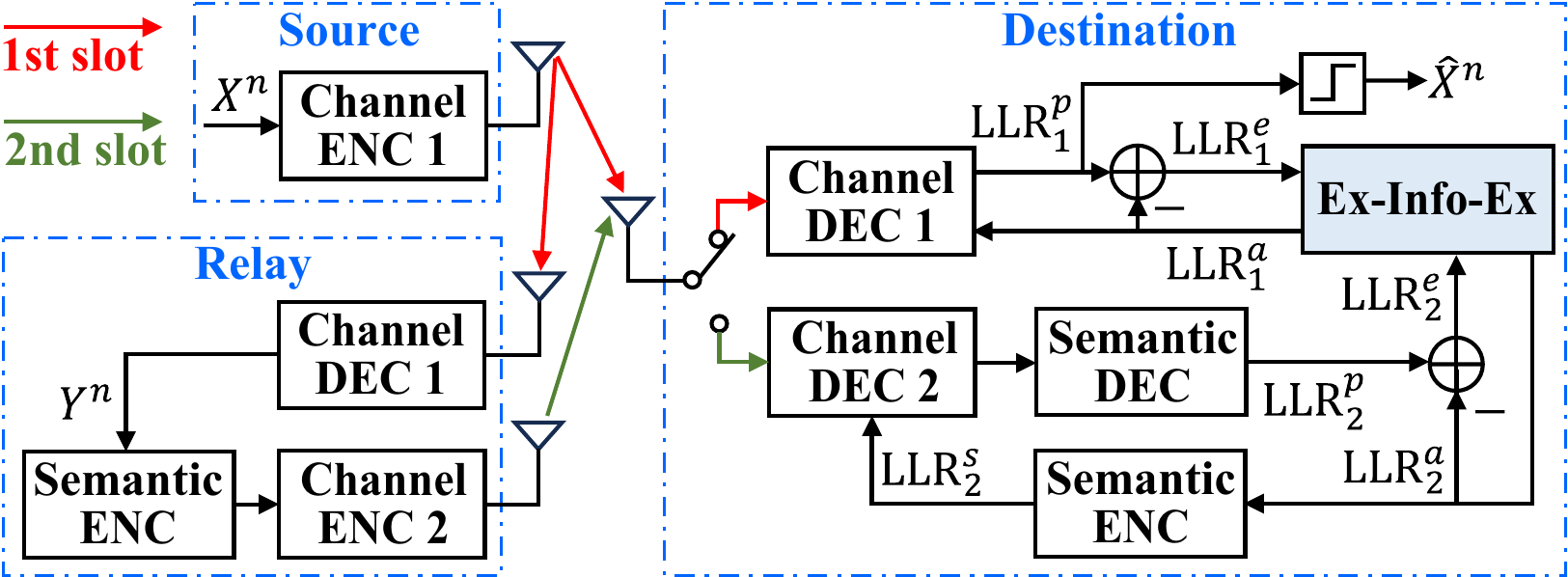}
\caption{The system structure of joint source-channel coding for SF relaying.}
\label{fig:ENC_DEC}
\end{figure}

\fref{fig:ENC_DEC} illustrates the general system structure of joint source-channel coding for SF relaying.
In practical systems, the quantization process may be needed if the media is transmitted in using symbols with finite alphabet.

Before the channel coding, the media information is quantized and represented by binary sequence with length $n$ in bits, as stated before.
After the channel decoding, the media information is recovered from the received binary sequence. 
Therefore, we focus on bit-wise joint decoding algorithms in this letter, and the design of symbol-wise joint decoding algorithms is left as the future work.

Hereafter, we utilize images as the examples for introducing the joint source-channel coding design of the media transmission with SF relaying.
One image pixel in each color channel is represented by 8 bits. 
For conciseness, the quantization process is omitted in description while implicitly used for converting the image pixel between analogue values and binary sequence.

At the source, information sequence $X^n$ is channel encoded by channel ENC 1, and then broadcast to the relay and the destination.
Then, the relay receives and decodes the source signals by channel DEC 1 to recover the source information sequence.
Due to the possible intra-link errors, the recovered sequence $Y^n$ may not be exactly the same as $X^n$.
Nevertheless, the relay continues encoding and transmitting the relay information to the destination.
The image feature is extracted by the semantic ENC, and further encoded by channel ENC 2 before transmitted to the destination.

At the destination, the received signals at the first slot are first decoded by channel DEC 1 to output the \emph{a posteriori} LLR ($\mathrm{LLR}_1^p$) of $X^n$.
The received signals of the second slot are successively decoded by channel DEC 2 and the semantic DEC to output the \emph{a posteriori} LLR ($\mathrm{LLR}_2^p$) of $Y^n$.

Then, the extrinsic LLR ($\mathrm{LLR}_i^e$) representing the decoding gain is calculated by 
$\mathrm{LLR}_i^e =  \mathrm{LLR}_i^p -\mathrm{LLR}_i^a, \textrm{ for } i\in \{1,2\},$
where $\mathrm{LLR}_i^a$ is the \emph{a priori} LLR, the initial value of which is all  set at $0$.
The extrinsic LLRs are then exchanged by the extrinsic information exchanger (Ex-Info-Ex).
Based on the correlation model of correlated sources \cite{garcia2005near}, the extrinsic LLRs of $X^n$ and $Y^n$ are exchanged by the LLR updating function $f_c(\cdot)$ \cite[Eq. (10)]{zhou2012great} to update the \emph{a priori} LLR. 

In the next round of decoding iteration, 
the updated $\mathrm{LLR}_1^a$ is directly utilized as the \emph{a priori} information for channel DEC 1. 
However, the updated $\mathrm{LLR}_2^a$ needs to be further encoded as the \emph{a priori} LLR ($\mathrm{LLR}_2^s$) of the semantic information, and then input into channel DEC 2.

Finally, if the iteration round reaches the maximum times, the latest $\mathrm{LLR}_1^p$ is utilized to reconstruct $\hat{X}^n$ by hard decision.

At the cost for reducing the link payload, SF requires relatively higher computational complexity than other relaying schemes, due mainly to the semantic encoder/decoder in the relay and the destination.
This algorithm presented in this letter  can be further extended to the system with multiple relays.
The example Python codes of the proposed SF relaying system are provided in \cite{SFgithub}.

\section{Performance Evaluation}\label{sec:Evaluation}

\subsection{Semantic Encoder/Decoder Structure}
Since the objective of this letter is to evaluate the performance gain of SF relaying, we design the structure of semantic encoder/decoder based on convolution and transpose convolution for simplicity.

\begin{table}[!t]
\renewcommand{\arraystretch}{1.1}
\tabcolsep=0.15cm
\caption{Neural Network Structure of the Semantic ENC/DEC}
\label{table:ENC_layers}
\centering
\begin{tabular}{c|c|c|c}
\hline
 \multicolumn{2}{c|}{\bfseries Layer type} & \bfseries Parameters & \bfseries Output shape \\
\hline
\multirow{4}{*}{ENC} & Input & None & $ 3 \times 96 \times 96$  \\
\cline{2-4}
& Conv2D & Filters: 16, Size: (2,2), Stride: 1 & $ 16 \times 95 \times 95$ \\
\cline{2-4}
& Conv2D & Filters: 16, Size: (3,3), Stride: 2 & $ 16 \times 47 \times 47$ \\
\cline{2-4} 
& Conv2D & Filters: 16, Size: (3,3), Stride: 2 & $ 16 \times 23 \times 23$ \\
\hline  
\multirow{4}{*}{DEC} & Input & None & $16 \times 23 \times 23$  \\
\cline{2-4}  
& TranConv2D & Filters: 16, Size: (3,3), Stride: 2 & $ 16 \times 47 \times 47$ \\
\cline{2-4}  
& TranConv2D & Filters: 16, Size: (3,3), Stride: 2 & $ 16 \times 95 \times 95$ \\
\cline{2-4} 
& TranConv2D & Filters: 3, Size: (2,2), Stride: 1 & $ 3 \times 96 \times 96$ \\
\hline  
\end{tabular}
\end{table}


In general, the design of neural network structures is affected by datasets.
The simulation images are randomly selected from the CIFAR-10 dataset \cite{CIFAR10}.
Thus, the neural network structures of the semantic ENC/DEC are designed as shown in Table \ref{table:ENC_layers}, where the padding is set at $0$ and omitted for all convolutional layers, and transpose convolutional layers.

For the semantic ENC, the input is a three-channel image with its width and height both being $96$ pixels. 
The $3 \times 96 \times 96$ pixel values are input into a two-dimensional (2D) convolutional layer (Conv2D) with $16$ filters, $(2,2)$-sized kernels, and a stride of $1$, to extract the image features into $16$ channels.
Subsequently, the features of $16$ channels are successively compressed by two concatenated Conv2Ds with $16$ filters, $(2,2)$-sized kernels, and a stride of $2$.
Therefore, the final output of the semantic ENC is the features of $16$ channels with the size in each channel being $23 \times 23$.
It is easy to calculate the compression rate of the semantic ENC as 
$(16 \times 23 \times 23)/(3 \times 96 \times 96) \approx 0.306$.

The semantic decoding process is the inverse of the semantic encoding process.
Therefore, for the semantic DEC, the input of the $(16 \times 23 \times 23)$-sized features is decompressed by two concatenated 2D transpose convolutional layers (TranConv2Ds) with $16$ filters, $(3,3)$-sized kernels, and a stride of $2$.
Finally, the $96 \times 96$ image pixels in the 3 channels are reconstructed by TranConv2D with $3$ filters, $(2,2)$-sized kernels, and a stride of $1$.

\subsection{Simulation Settings}
In the simulations, we assume \gls{LoS} components dominate the channels for simplicity.
Compared to the source, we assume the relay is closer to the destination, and hence the \gls{SNR} $\gamma_2$ in the \gls{RD} link is relatively higher than the SNR $\gamma_1$ in the \gls{SD} link.
We simply set $\gamma_2=20$ dB which is sufficiently large for losslessly transmitting the semantic information, while $\gamma_1$ varies from $-5$ dB to $9$ dB.
\Gls{AWGN} channels are adopted for the \gls{SD} and \gls{RD} links.
For the \gls{SR} link, we utilize the crossover probability $\rho$ between the quantized bits $X^n$ and $Y^n$ to represent the channel conditions for simplicity.

The \gls{LDPC} codes \cite{gallager1962low} are utilized as the channel codes, with the codeword length being set at $900$. 
There are $3$ bits in the same parity-check equation, and each bit is associated with $2$ parity-check equations.
To satisfy the LDPC codeword length, the information bits are divide into groups, and the last group is padded by zero bits.
The maximum number of the local decoding iteration for LDPC codes, and that of the global decoding iteration are set at $1$ and $7$, respectively. 
Before the first global iteration, we need one round of initial LDPC decoding for obtaining non-zero LLRs, and hence the ensemble number of LDPC decoding iterations is equal to $8$.

The loss function for training the semantic neural network is defined as 
$Loss=\min [ \alpha \mathcal{F}_{\mathrm{MSE}}(\boldsymbol{X} ,  \boldsymbol{\hat{X}} )+ \beta \mathcal{F}_{\mathrm{CE}}( \boldsymbol{Z} ,  \boldsymbol{\hat{Z}} ) ],$	
where $ \boldsymbol{X} $ and $ \boldsymbol{\hat{X}} $ are the input and output image matrix, respectively. 
$ \boldsymbol{Z} $ and $ \boldsymbol{\hat{Z}} $ are matrices for distinguishing different categories with the same batch size as $ \boldsymbol{X} $ and $ \boldsymbol{\hat{X}} $. 
We use $\mathcal{F}_{\mathrm{MSE}}$ and $\mathcal{F}_{\mathrm{CE}}$ to evaluate mean square error and cross entropy, respectively. 
The hyper-parameters $\alpha$ and $\beta$ are used to balance the proportion of $\mathcal{F}_{\mathrm{MSE}}$ and $\mathcal{F}_{\mathrm{CE}}$. 
The initial parameters of the semantic neural network are set as follows. 
The hyper-parameters $\alpha$ and $\beta$ are set at $1.5$ and $0.56$, respectively. 
The Adam optimizer \cite{kingma2014adam} is adopted with a learning rate of $0.001$.
Moreover, we employ a batch size of $64$ and $200$ epochs for training. 

\subsection{Simulation Results}
%
%
%

\begin{figure}[!t]
\centering \includegraphics[width=\figwidthS]{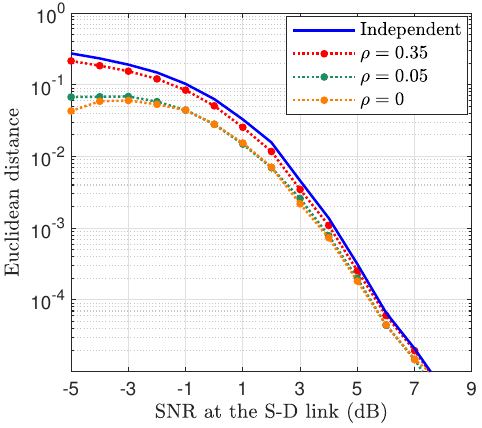}
\caption{Euclidean distance with diverse $\rho$.}
\label{fig:ED}
\end{figure}

To quantify the similarity between the original and reconstructed images, we use the \gls{ED} as a reliable metric.
Due to the page limit, the comparison with other relaying systems is left as the future work.
\fref{fig:ED} plots the \gls{ED} curves for diverse $\rho$.
Obviously, the joint decoding with SF relaying always outperforms the independent decoding, which verifies the effectiveness of SF.
Another observation that justifies our intuition is that, the ED reduces as the quality of the relay information increases, i.e., smaller $\rho$.
Notice that the performance gain decreases as the SNR of the S-D link increases.
This is because the LDPC codes have a greater capability for increasing coding gains and hence the independent LDPC decoding has already corrected most of the errors, when the \gls{SD} link SNR becomes larger.
In particular, the ED of the independent decoding decreases to $0$ when $\gamma_1 \ge 8$ dB.


\begin{figure}[!th]
\centering 
\subfigure[Example with $\rho=0$.]{
\includegraphics[width=1.05in]{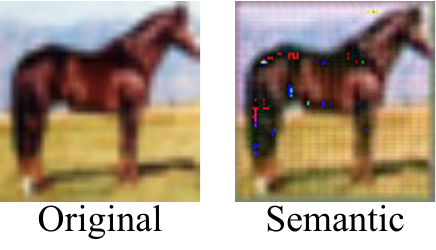}
\label{fig:original}
}

\subfigure[$\gamma_1=-5$ dB, $\rho=0.1$.]{
\includegraphics[width=\figwidthW]{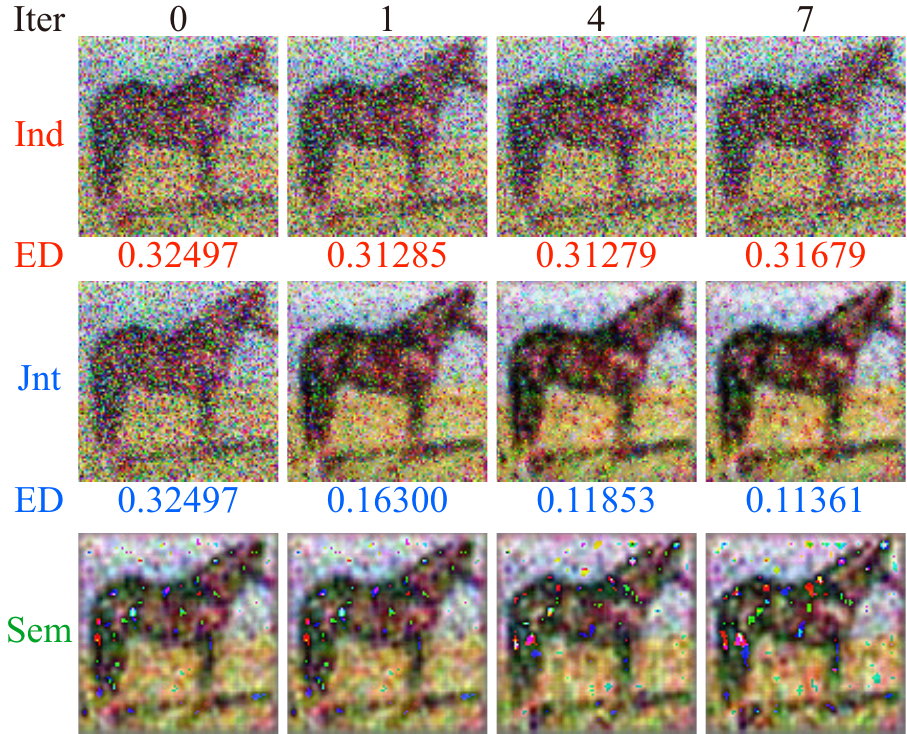}
\label{fig:snr-5-rho0.1}
}
\subfigure[$\gamma_1=-5$ dB, $\rho=0.35$.]{
\includegraphics[width=\figwidthW]{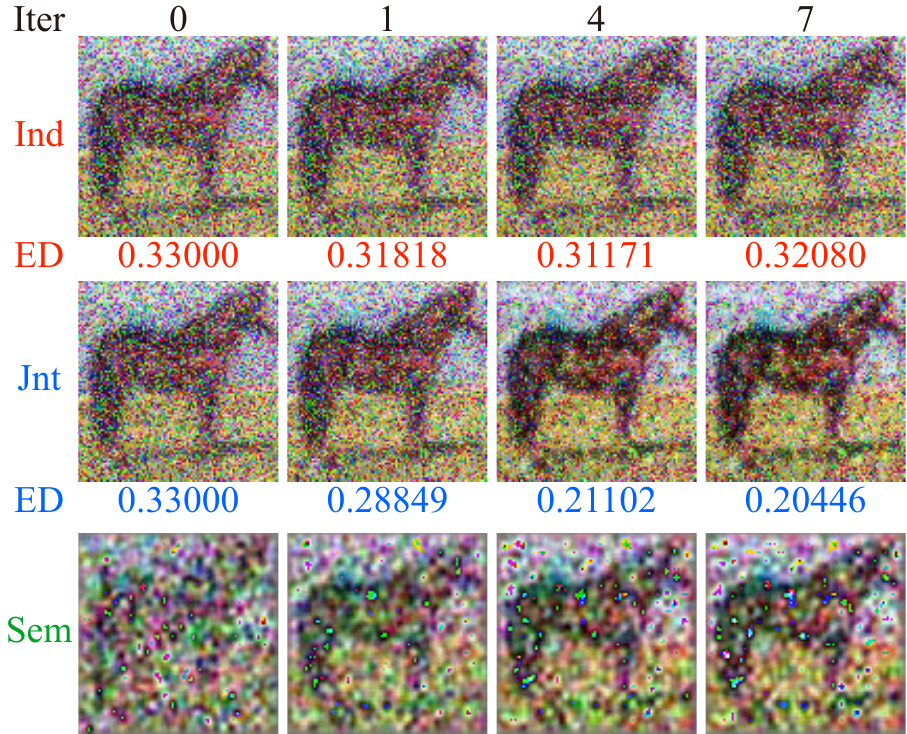}
\label{fig:snr-5-rho0.35}
}
\subfigure[$\gamma_1=0$ dB, $\rho=0.35$.]{
\includegraphics[width=\figwidthW]{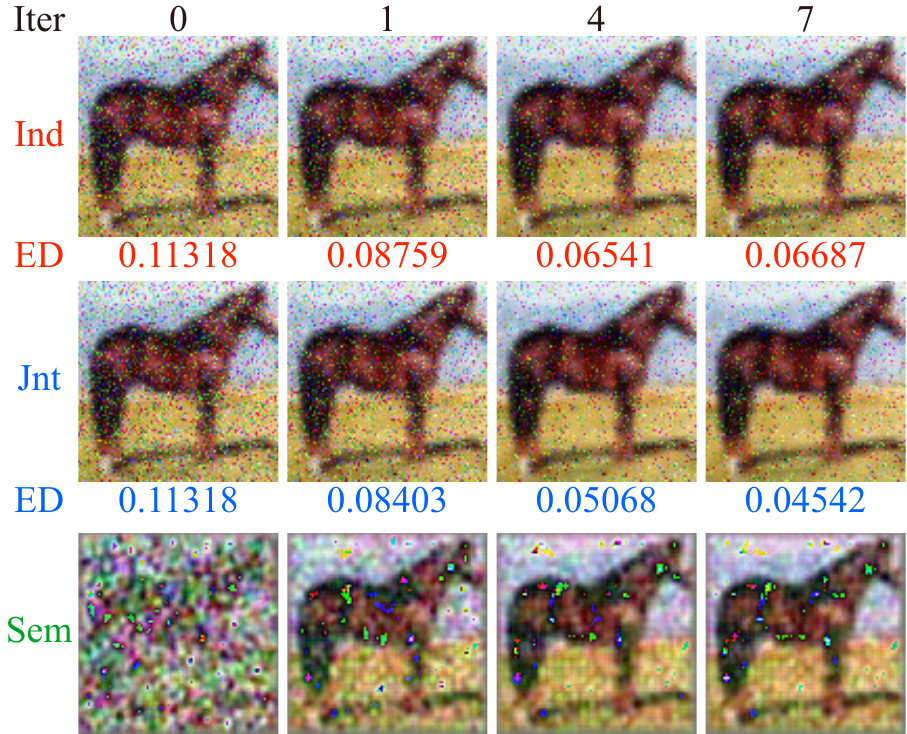}
\label{fig:snr0-rho0.35}
}
\caption{Comparison of the reconstructed image qualities.}
\label{fig:image_quality}
\end{figure}

We utilize the original and error-free semantic images shown in \fref{fig:original} as an example for displaying the reconstructed images.
Figs. \ref{fig:snr-5-rho0.1}-\ref{fig:snr0-rho0.35} depicts the reconstructed images.
Clearly, the image quality increases as the round of global iterations (Iter) increases.
It is found that no more image quality improvement can be achieved when Iter changes from $4$ to $7$, and hence Itr$=7$ is enough in this system setup.

In \fref{fig:snr-5-rho0.1}, it is clearly found that the images of joint (Jnt) decoding have a higher visual quality than the images of independent (Ind) decoding.
The images of semantic (Sem) decoding also becomes clearer by exchanging the extrinsic information obtained via the \gls{SD} link.

\fref{fig:snr-5-rho0.35} shows the results in very bad channel conditions.
Surprisingly, although the initial semantic image is unresolvable, it exhibits a relatively clear shape of the object after joint decoding.
Due to the exchange of the extrinsic information, the image quality of joint decoding also improves dramatically.

By comparing Figs. \ref{fig:snr-5-rho0.35} and \ref{fig:snr0-rho0.35}, we can conclude that the image quality is higher in better channel conditions.
Moreover, the semantic image is recovered very similar to the case with $\rho=0$ shown in \fref{fig:original}.

\section{Conclusion}\label{sec:Conclusion}
We have proposed a novel concept of SF relaying, which is suitable for 6G media transmissions and adaptively accommodates various types of information.
The principle of SF relaying has been explained in detail, and its achievable rate constraints for lossless recovery at the destination have been analyzed.
In addition, we have designed a joint source-channel coding scheme for SF relaying, and further implemented the SF relaying technique in image transmission simulations.
The simulation results indicate that the SF relaying system can adequately eliminate the effect of intra-link errors by utilizing the semantic decoder, reduce the payload in the \gls{RD} link, and achieve lossless transmissions in the worse channel conditions.
%

\bibliographystyle{IEEEtran}
\bibliography{myreference}

\end{document}